\documentclass[pra,a4paper,twocolumn,showpacs]{revtex4}
\usepackage{amsmath,amsfonts,graphicx,color}
\usepackage{amssymb}
\def\>{\rangle}
\def\<{\langle}
\def\per{\text{per}}

\def \be{\begin{equation}}
\def \ee{\end{equation}}
\def \beq{\begin{equation}}
\def \eeq{\end{equation}}
\def \bea{\begin{eqnarray}}
\def \eea{\end{eqnarray}}

\begin{document}

\title{A simple encoding of a quantum circuit amplitude as a matrix permanent}
\author{Terry Rudolph}%
\affiliation{Optics Section, Blackett Laboratory, Imperial College
London, London SW7 2BW, United Kingdom}%
\affiliation{Institute for Mathematical Sciences, Imperial College London, London SW7 2PG,
United Kingdom}%

\date{\today}

\begin{abstract}

A simple construction is presented which allows computing the transition amplitude of a quantum circuit to be encoded as computing the permanent of a matrix which is of size proportional to the number of quantum gates in the circuit. This opens up some interesting classical monte-carlo algorithms for approximating quantum circuits.

\end{abstract}

\maketitle

In a recent article \cite{loebl} Loebl and Moffatt gave a method for expressing the computation of the Jones polynomial of a braid in terms of a matrix permanent. Although computing permanents is believed difficult (\#P-complete in the language of complexity theory), there exist probabilistic algorithms \cite{godsil} which sample the permanent, and this suggests some interesting new classical algorithms for estimating the output amplitudes of quantum circuits because evaluating the Jones polynomial at certain roots of unity is BQP-complete \cite{freedman}. The route to encoding a quantum circuit as the Jones polynomial of a knot, and then as a matrix permanent, is somewhat complicated - the purpose of this article is to present a simpler construction.

We restrict to quantum circuits built from Toffoli and Hadamard gates, which are universal \cite{shih}. We rely heavily on the construction of Dawson et al \cite{dawson}. There is it shown how the transition amplitude for such a quantum circuit is equivalent to counting the number of solutions of a GF(2) (i.e. XOR-AND) polynomial over some binary valued variables. More precisely, the results of \cite{dawson} imply the following: Given a quantum circuit $U$ and in/output computational basis states $|\text{in}\>,|\text{out}\>$ the amplitude $\<\text{out}|U|\text{in}\>$ can be expressed as the \emph{difference} in the number of solutions to a GF(2) polynomial over (roughly) as many boolean variables as there are Hadamard gates in the circuit. It is perhaps easiest to explain the construction using an example, such as in Figure 1. The $a_i,b_i,..$ etc are boolean variables, which we imagine travelling along the qubit lines. Every time the qubit goes through a Hadamard gate we create a new such variable, and  whenever a variable $z_i$ travels through the target of a Toffoli gate we replace it by $z_i\oplus x_iy_i$ where $x_i,y_i$ are the variables at the control lines of the Toffoli gate, as indicated.

\begin{figure}
  \includegraphics*[width=9.0cm, bb=0.5cm 4.5cm 27.3cm 18.7cm]{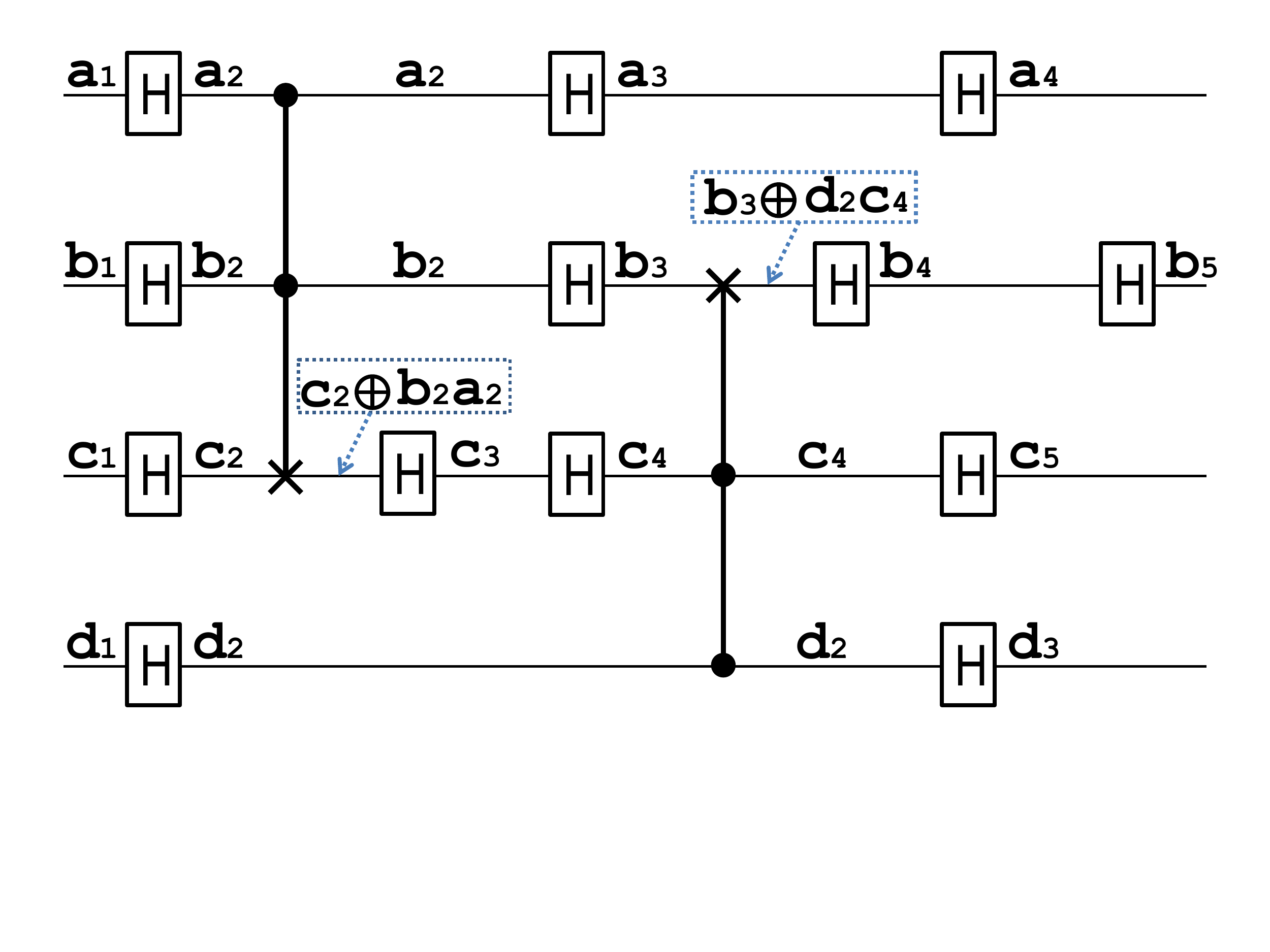}
\caption{Mapping from a standard Toffoli-Hadamard circuit to counting solutions of a GF(2) polynomial}
\label{circuitexample}
\end{figure}

Having labelled the circuit with these variables, we then create the function $f(x)$ by taking the sum (mod 2) of the product of every pair of variables on either side of a Hadamard gate. For the example of Fig.~\ref{circuitexample} we obtain:
\begin{eqnarray*}
f(x)&=&a_1a_2\oplus a_2a_3 \oplus a_3a_4 \oplus \\
&& b_1b_2 \oplus b_2b_3 \oplus (b_3 \oplus d_2c_4)b_4 \oplus b_4b_5\oplus\\
&& c_1c_2 \oplus (c_2 \oplus b_2a_2)c_3 \oplus c_3c_4 \oplus c_4c_5\oplus \\
&&d_1 d_2 \oplus d_2d_3
\end{eqnarray*}
If we are interested in, for example, the amplitude $\<0011|U|0000\>$ we then fix the input and output variables of $f$ accordingly: in this case we would set $a_1=b_1=c_1=d_1=a_4=b_5=0$, $c_5=d_3=1$ and $f$ simplifies to:
\begin{eqnarray*}
f(x)&=& a_2a_3 \oplus b_2b_3 \oplus b_3b_4 \oplus d_2c_4b_4 \oplus c_2c_3 \oplus b_2a_2c_3\oplus \\
&& c_3c_4 \oplus c_4 \oplus d_2
\end{eqnarray*}

What is shown in \cite{dawson} is that given a function constructed in this way, one has:
\begin{equation}\label{qamp}
\<\text{out}|U|\text{in}\>=\frac{\#_0-\#_1}{\sqrt{2}^{h}}.
\end{equation}
Here $\#_0,\#_1$, denote the number of solutions to the equation $f(x)=0$, $f(x)=1$ respectively, and $h$ denotes the number of Hadamard gates in the circuit. Note that $\#_0+\#_1=2^{v}$ where $v$ is the number of variables in the function $f$ once the input and output qubit values have been fixed. If there are $q$ qubits in the circuit then  $v=h-q$.

There are several other points to note in terms of the construction of $f$. Firstly it will be convenient to assume that every variable goes through at most one Toffoli gate - this can be arranged by inserting double Hadamard (i.e. identity) gates where necessary. This should also be done at the final outputs to the quantum circuit. Doing so ensures that the function $f$ has the following properties (i) it is (monotone) cubic (ii) every variable appears in at most one cubic clause and two quadratic clauses.

\begin{figure}
  \includegraphics*[width=9.0cm, bb=0.2cm 9.5cm 27.5cm 19cm]{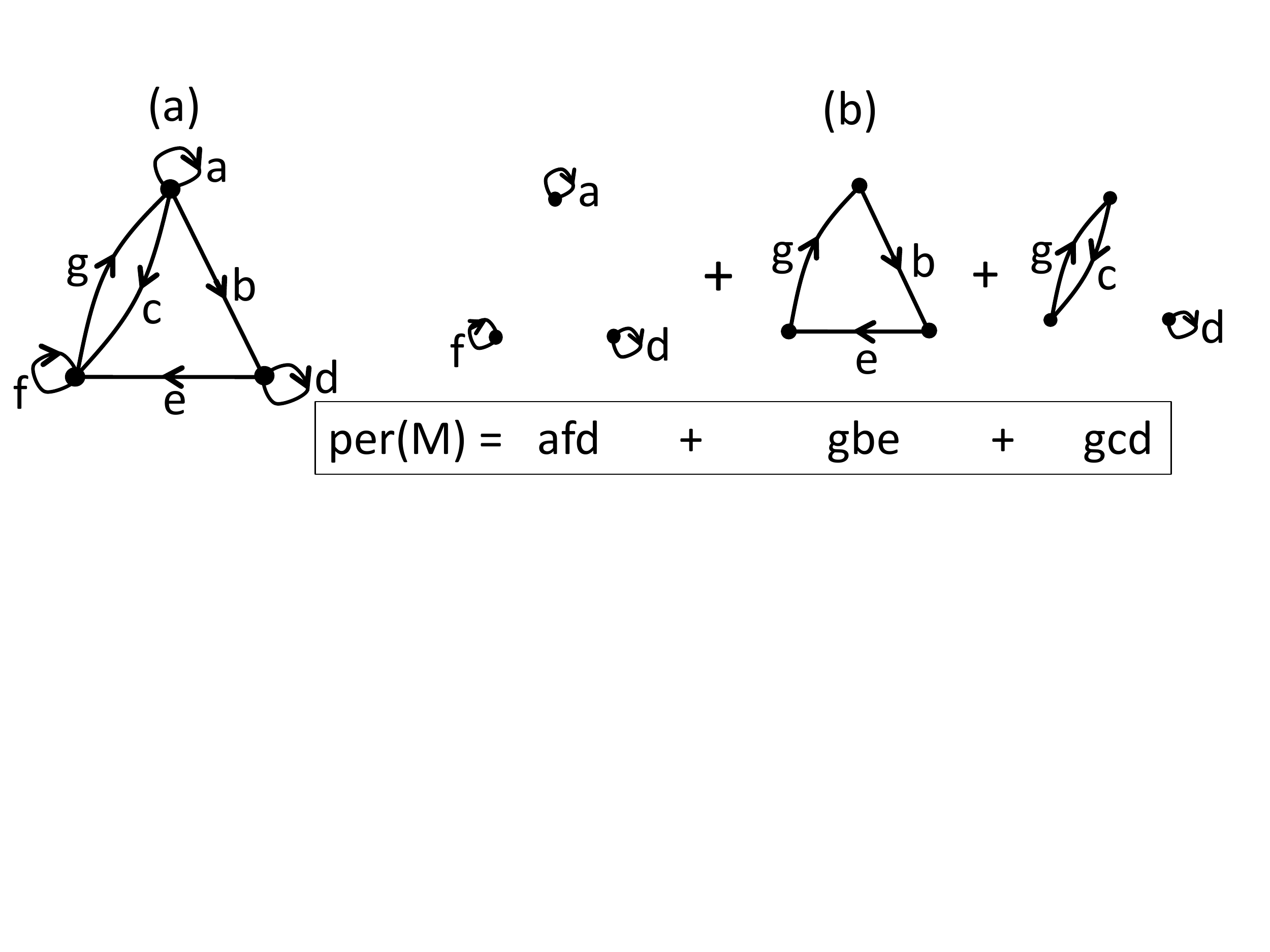}
\caption[]{The permanent of the matrix $M=\begin{pmatrix} a & b & c \\ 0 & d & e \\ g & 0 & f \end{pmatrix}$ is the sum of the weighted cycle covers of the associated graph.}
\label{cyclecovers}
\end{figure}

Now counting solutions to a general GF(2) polynomial is a $\#P$-complete problem \cite{karpinski}. That is, it has the same complexity as computing the permanent of a matrix - the prototypical $\#P$ problem - as was famously proven by Valiant in 1979 \cite{valiant}. So we know that in principle we can map between these problems and find some matrices $M_0$, $M_1$ such that $\per(M_0)=\#_0$, $\per(M_1)=\#_1$ and then  \[\<\text{out}|U|\text{in}\>=\frac{\per(M_0)-\per(M_1)}{\sqrt{2}^{h}}.\] However; the actual mapping between these problems is not particularly simple or economical. In addition Valiant's construction of the matrix to count solutions of a satisfiability problem is also not particularly economical.

The purpose of this article is to present a very simple, direct and economical construction relating quantum computing to evaluating a matrix permanent, which is also considerably more efficient than following the preceding route. Moreover; instead of expressing the solution to the problem as the difference in two matrix permanents, we will construct a single matrix/graph $G$ such that
\begin{equation}\label{myamp}
\<\text{out}|U|\text{in}\>=\frac{\per(G)}{\sqrt{2}^{h}}.
\end{equation}

The route to finding $G$ uses some of the same tricks as in Valiant's proof. As this paper is intended to also be accessible for physicists possibly unfamiliar with Valiant's result, we will try and make the presentation as self-contained as possible.

Any $n\times n$ matrix can be considered the weighted adjacency matrix for a weighted graph on $n$ vertices, where the weight on the edge between vertices $i$ and $j$ is simply the $(i,j)$'th element of the matrix. The permanent of a matrix, formally defined by \[per(M)=\sum_{\pi\in S_n}\prod_i M_{i,\pi(i)}\] with $S_n$ the symmetric group on $n$ symbols,  is then graphically equivalent to the sum total of the weighted cycle covers of the graph: A cycle in a graph is a closed path, a cycle cover is a set of cycles for which each vertex belongs to one and only one cycle. The weight of a cycle cover is the product of the weights on the edges involved in that particular cycle cover - so the permanent is the sum of all such weights. An example is provided in Fig.~\ref{cyclecovers}. A brief summary of how the permanent arises in some physical considerations can be found in \cite{simone}.

\begin{figure}
  \includegraphics*[width=9.0cm, bb=4.5cm 2.5cm 21cm 15cm]{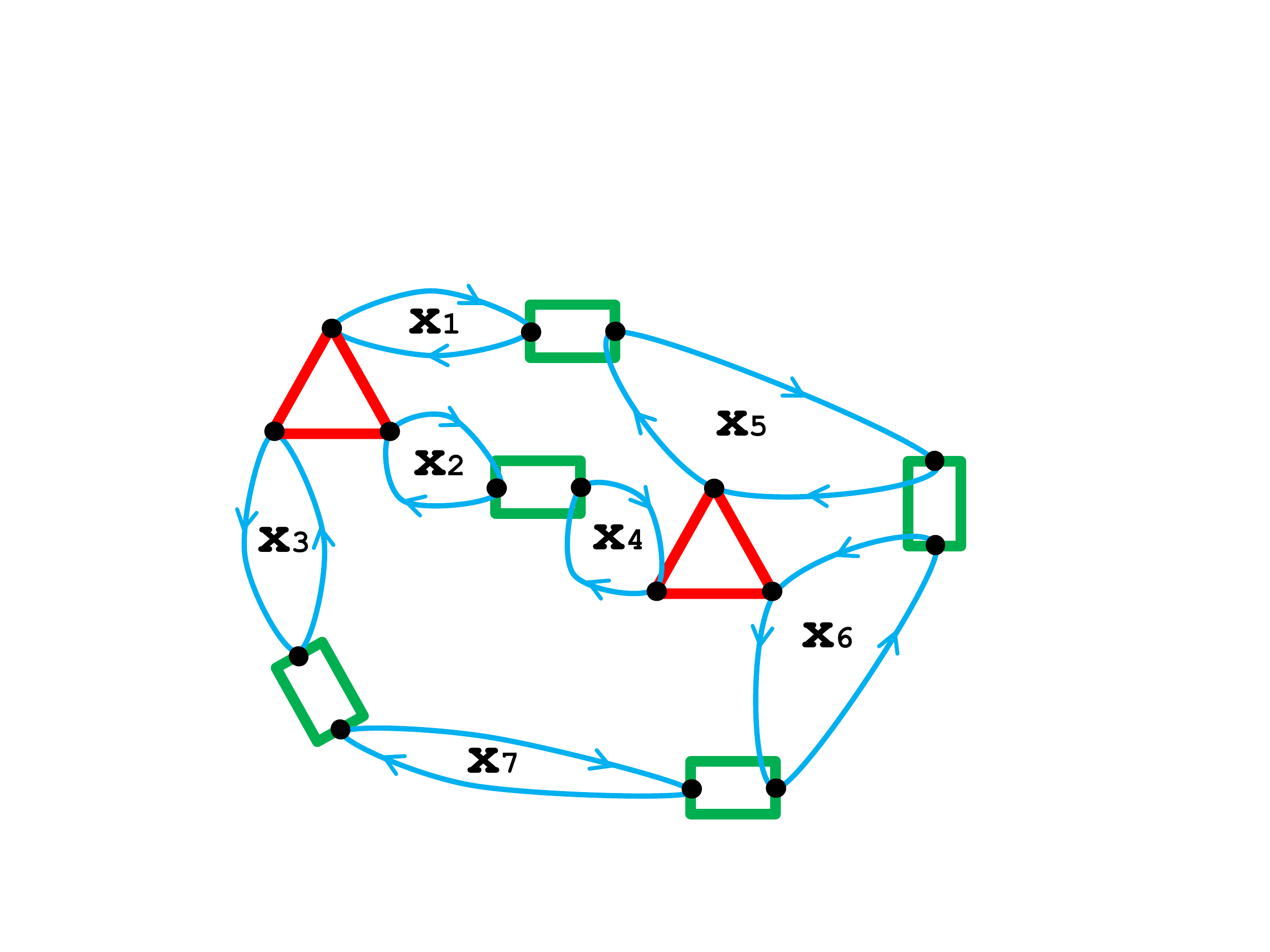}
\caption{(Color online) A big-picture view of the construction. The external edges form loops through the graph gadgets, and each such cycle is associated with one particular Boolean variable $x_i$. If the cycle \emph{is} traversed in a particular cycle cover then that corresponds to setting that particular variable to 0. Conversely, if the particular cycle \emph{is not} traversed then this corresponds to the associated variable having a value of 1. The graph gadgets have 2 or 3 vertices connecting to external edges according to whether they are gadgets for a quadratic or cubic clause. This graph would correspond to the polynomial $x_1x_2x_3\oplus x_1x_5 \oplus x_4x_5x_6\oplus x_5x_6 \oplus x_2x_4 \oplus  x_6x_7\oplus x_3x_7$}
\label{bigpicture}
\end{figure}

Let us first give an overall view of the construction. We will be constructing a graph in such a way that the presence/absence of one particular cycle in any given cycle cover corresponds to whether a particular boolean variable $x_i$ associated with this cycle is 0 or 1. We will use the convention that if the particular cycle is present in the cycle cover then this matches the variable assignment $x_i=0$, if it is not then $x_i=1$. Not all cycles within the graph will correspond to variable assignments - the ones which do we term external cycles. In figures the ``external edges'' which can make up such cycles will be colored in blue (to aid the eye only - there is no mathematical difference between these edges and other edges in the graph). The overall graph will consist of some ``graph gadgets'' (small subgraphs) connected by external edges. An example is given in Fig.~\ref{bigpicture}. Each of the gadgets corresponds to a clause - in the figure we show only the vertices of the gadget which connect to external edges. The blue external edges form loops around two or three of the graph gadgets according to whether the variable appears in two or three clauses, and obviously they loop through a clause gadget with their corresponding partners of that clause looping through the other vertices of the gadget.

\begin{figure}
  \includegraphics*[width=8.0cm, bb=6.5cm 3cm 19cm 17.5cm]{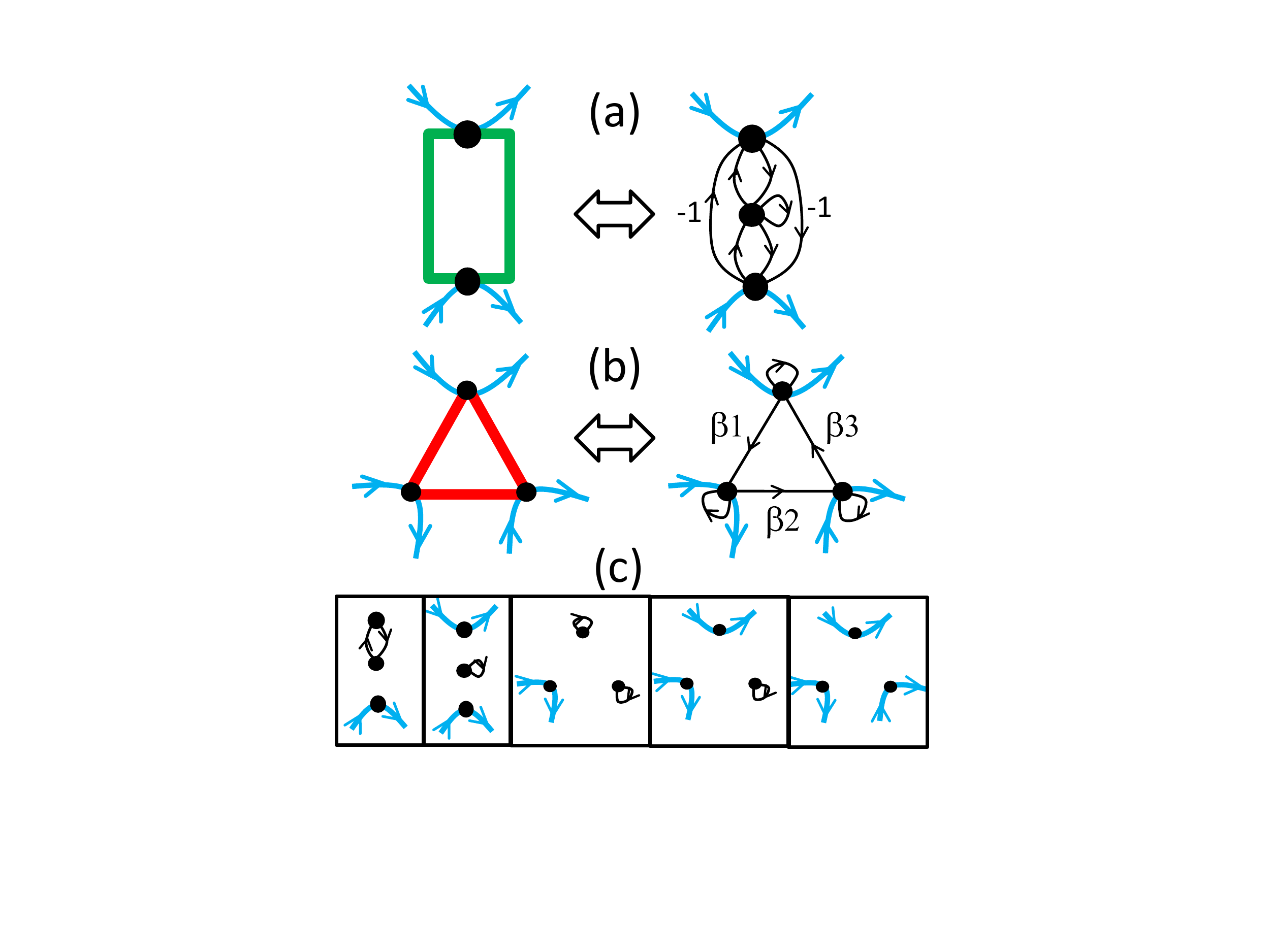}
\caption{(a) The inner workings of the quadratic clause graph gadget. The weight on any edge is 1 unless otherwise indicated. (b) The inner workings of the cubic clause gadget. The weights $\beta_i$ need only satisfy $\beta_1\beta_2\beta_3=-2$, which can be achieved by setting $\beta_1=-2,\beta_2=\beta_3=1$ if a graph with integer weights is desired. (c) The various ways in which the gadgets can be involved in a cycle cover with external edges. These are the cases when some of the associated boolean variables in the clause are equal to 0 and, as can be seen, these cases all contribute a weight of +1 to the cycle cover. When no external edges are incident on the gadget the weight it contributes must be -1, as discussed in the text.}
\label{bothgadgets}
\end{figure}

Now as we compute the sum of the weighted cycle covers of the graph (ie the permanent of the associated matrix) each cycle cover in the sum corresponds to a particular assignment of values to the Boolean variables - i.e. it will have a particular set of external cycles traversed, setting those variables to a value 0. The graph gadgets will be designed so that if \emph{none} of the external edges connected to that gadget are traversed - corresponding to all of the variables in that clause being equal to 1 - then the weight which that gadget contributes to the particular cycle cover is $-1$. In all other cases the weight contributed by that gadget will be $+1$. Recall that the weight of any given cycle cover is the \emph{product} of the weights over all cycles in the cover. So for a fixed cycle cover (corresponding to a fixed assignment to the boolean variables) the total weight will be $+1$ or $-1$ according to whether an even or an odd number of clauses are satisfied by that particular assignment. Assuming w.l.o.g. an even number of clauses in total, this in turn means that the weight of the particular cycle cover is +1 if $f(x)=0$ and $-1$ if $f(x)=1$. As we sum over all weighted cycle covers we automatically are calculating the difference in the number of solutions of $f(x)=0$ to $f(x)=1$, which is precisely what we need by Eq.~{qamp}.

The inner working of  graph gadgets which act in the desired manner are shown in Figs.~\ref{quadraticgadget}. If in some cycle cover no external edges are incident on the gadget then its contribution will be $-1$, which can be readily verified by computing the permanents of their adjacency matrics:  $\begin{pmatrix}0 & -1 & 1 \\-1 & 0 & 1 \\1 & 1 & 1 \end{pmatrix}$ and $\begin{pmatrix}1 & \beta_1 & 0 \\0 & 1& \beta_2 \\\beta_3 & 0 & 1 \end{pmatrix}$. If one, two or three external edges are incident on the gadgets then the contribution to the cycle cover has weight +1, this is depicted in Fig.~\ref{bothgadgets}(c).

\begin{figure}
  \includegraphics*[width=9.0cm, bb=7.3cm 6.7cm 22.2cm 14cm]{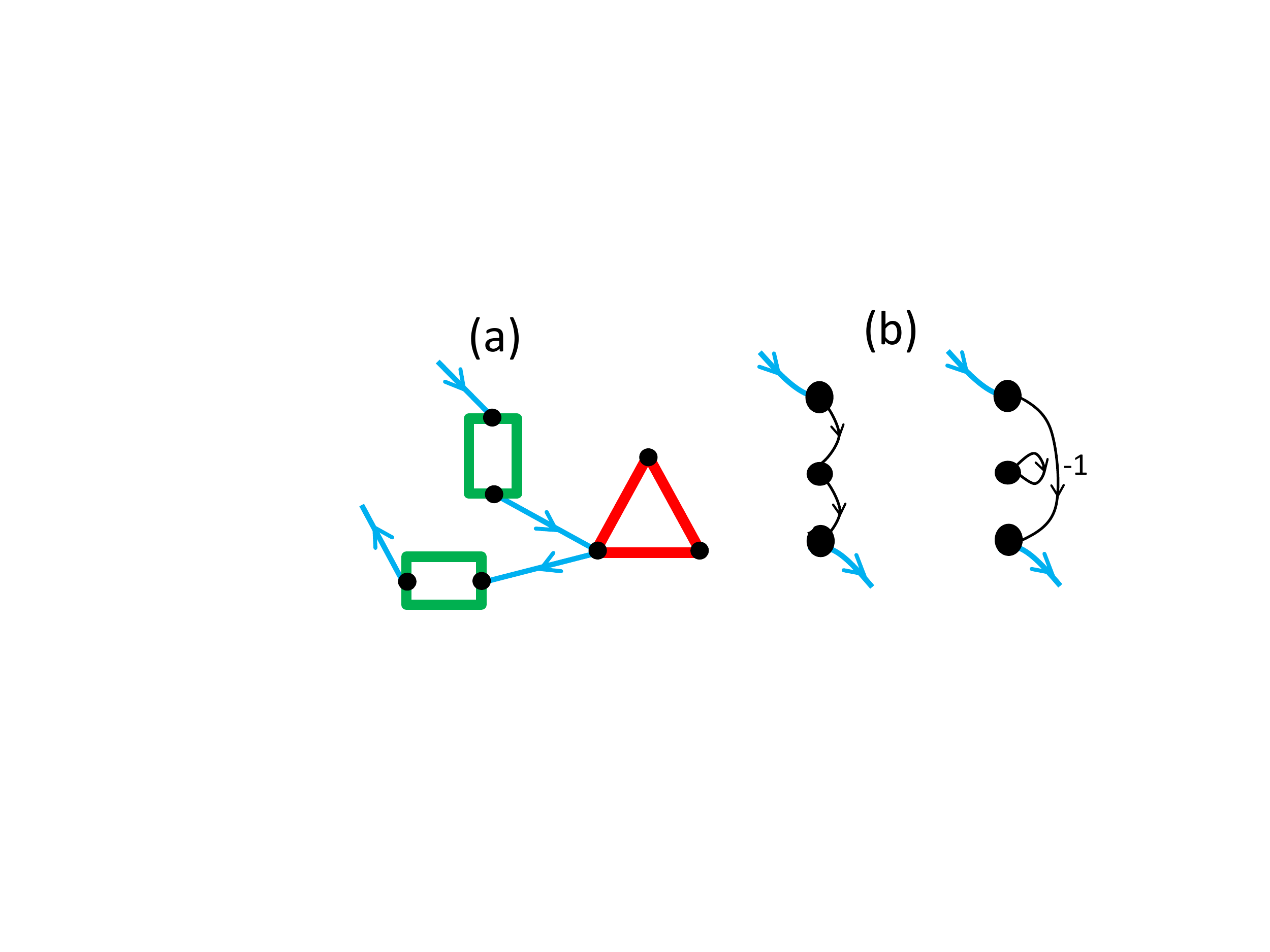}
\caption{(a) The sort of cycle covers we need to avoid cycles which only partially traverse an external cycle, as such setting the associated variable to 0 in one clause and 1 in another. (b) The inner workings of the quadratic clause gadget which ensure that any external edge must exit by the same vertex it entered. The two depicted contributions to the cycle cover have opposite signs and cause the necessary cancelation.}
\label{avoidingbadcycles}
\end{figure}

There is one potential problem which has not been addressed. What is to stop a particular cycle cover involving only \emph{part} of an external cycle corresponding to some given variable. Why, for example, do we not get screwed up by cycle covers which, say, enter at one vertex of the graph gadget but leave at a different one? A figurative picture of such an undesirable type of cycle is given in Fig.~\ref{avoidingbadcycles}(a).

The possibility of such problematic cycles is ruled out by the internal workings of the quadratic clause graph gadget. This is shown in Fig.~\ref{avoidingbadcycles}(b). Any cycle cover which enters the gadget along one external edge and tries to leave out via the external edge on the other side of the gadget has two possible paths for doing so. These paths pick up opposite signs, and so when summed over contribute 0 to the total. The process is somewhat reminiscent of mach-zender interferometry! Note that we did not need to design the cubic graph gadget to have the same property. This is because in the formulation we have chosen any variable appears in only one cubic clause, and it must then also appear in two quadratic clauses. The quadratic clause gadgets suffice to ``force'' an external edge which is incident into the cubic clause gadget to leave via the same vertex it entered.

\begin{figure}
  \includegraphics*[width=9.0cm, bb=0.6cm 4.5cm 25cm 19cm]{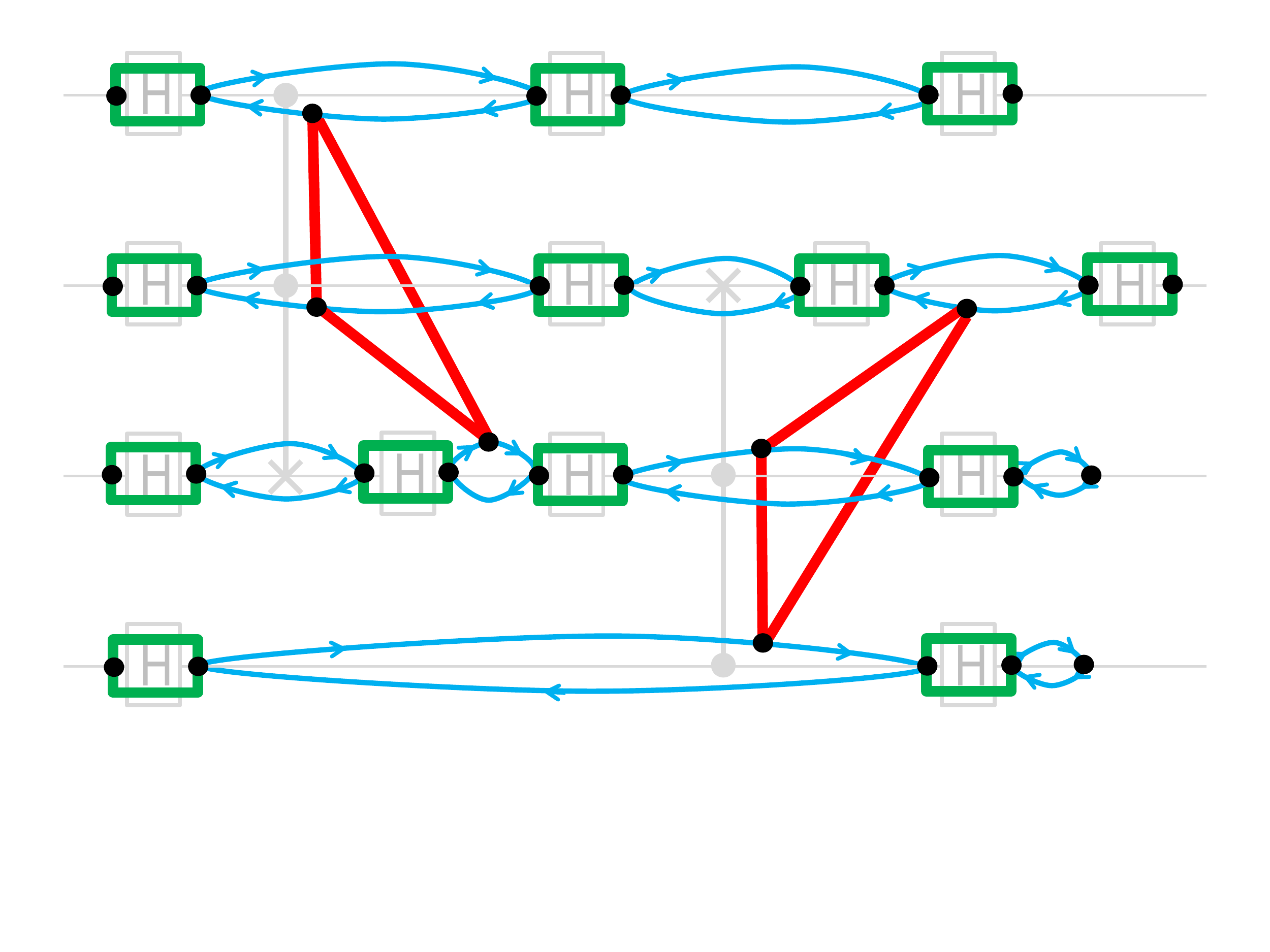}
\caption{Putting everything together - how to draw the final graph $G$ over the top of the associated circuit. Note that it is the variable on the target line of a Toffoli gate which is created by the Hadamard that acts \emph{after} the Toffoli which actually is involved in the associated cubic clause. This graph would be computing the transition amplitude with $|\text{in}\>=|1111\>$ and $|\text{out}\>=|1100\>$. }
\label{circuitandgraph}
\end{figure}

In terms of the basic construction the final thing to mention is that it is simple to force the values at the boundaries (the input/output to the circuit) to be 1 or 0. This is done either by simply not connecting any external edges into the associated gadget (setting the variable to 1), or by forcing an external edge through the gadget by having that edge also loop through a vertex which has no ``self loop'' (setting the variable to 0). An example of this can be seen in Fig.~\ref{circuitandgraph} where the input qubits are all fixed to have value 1, and the top two qubits have value 1 at the output while the bottom two qubits are set to the value of 0 at the output\footnote{In practise we can make things slightly more economical by removing some of these redundant vertices and using the fact that a suitable clause gadget for a clause consisting of a single variable is simply a single vertex with a self-loop of weight -1.}.

The overall construction can be naturally laid out by drawing the graph directly on top of the circuit diagram. This is illustrated in Fig.~\ref{circuitandgraph} for the same circuit of Fig.~\ref{circuitexample}.

Note that the number of vertices in the graph $G$ we associate to an given circuit is basically 3 times the number of gates in the circuit. Let us denote this number of vertices as $m$. We have that 
\[
\<\text{out}|U|\text{in}\>=\frac{\per(G)}{\sqrt{2}^{h}}=\per\left(\frac{G}{\sqrt{2}^{h/m}}\right).
\] If it were the case that $\|\frac{G}{\sqrt{2}^{h/m}}\|<1$ then the results of \cite{gurvits} imply there would exist an efficient classical algorithm to simulate this quantum circuit. 

\begin{acknowledgments}
We acknowledge the support of the EPSRC and helpful comments by S. Severini.
\end{acknowledgments}


\begin{thebibliography}{99}

\bibitem{loebl}  M. Loebl and I. Moffatt, arxiv:0705.4548.

\bibitem{godsil} C.D. Godsil and I. Gutman in Algebraic Methods in Graph Theory, Vol. I, II,(Szeged, 1978), North-Holland, Amsterdam-New
York, 1981; N. Karmarkar, R. Karp, R. Lipton, L. Lovasz, M. Luby, SIAM J. Comput. \textbf{22}, 284, (1993); A. Barvinok, Ran. Struct. Algor. \textbf{14}, 29–61 (1999); 
M. Jerrum, A. Sinclair and E. Vigoda, Journal of the ACM, \textbf{51} 671 - 697 (2004).  


\bibitem{freedman} M. Freedman, A. Kitaev, M. Larsen and Z. Wang, Bull. Amer. Math. Soc. \textbf{40}, 31 (2003);
M. Bordewich, M. Freedman, L. Lovasz and D.Welsh, Combinatorics, Probability and Computing \textbf{14}, 737, (2005);
D. Aharonov, V. Jones and Z. Landau, Proceedings STOC 06, 427 (2006).

\bibitem{shih} Y. Shi, Quantum
Information and Computation, \textbf{3}, 84 (2003)2003; D. Aharonov, arXiv:quant-ph/0301040, 2003.

\bibitem{dawson} C. M. Dawson et al, Quantum Information and Computation, \textbf{5}, 102 (2005); arxiv:quant-ph/0408129

\bibitem{karpinski} A. Ehrenfeucht and M. Karpinski,  The Computational Complexity of (XOR, AND)-Counting Problems, ICSI Technical Report TR-90-033, July 1990.

\bibitem{valiant} L. G. Valiant, Theoret. Comput. Sci. \textbf{8}, 189 (1979)

\bibitem{simone} T.-C. Wei and S. Severini, arxiv:0905.0012


\bibitem{gurvits}  L. Gurvits, On the Complexity of Mixed Discriminants and Related Problems, Mathematical Foundations of Computer Science 2005, Springer Berlin / Heidelberg
Vol. 3618, (2005). 
\end{thebibliography}
\end{document}